# Construction of an ESR-STM for Single Molecular Based Magnets anchored at Surfaces

*Paolo Messina[1,2], Paolo Sigalotti[2], Lorenzo Lenci[3], Stefano Prato[2], Matteo Mannini[1] Paolo Pittana[4], Dante Gatteschi[1]

1) INSTM UDR Firenze, Sesto Fiorentino, 50119, Italy 2)APEResearch Area Science Park, Trieste, 34012, Italy 3) IPCF Area Della Ricerca CNR, Pisa, 56124, Italy , Sincrotrone Trieste, Area Science Park, Trieste, 34012, Italy

*ABSTRACT* — **Reading and manipulating the spin status of single magnetic molecules is of paramount importance both for applicative and fundamental purposes. The possibility to combine Electron Spin Resonance (ESR) and Scanning Tunnelling Microscopy (STM) has been explored one decade ago. A few experiments have raised the question whether or not an EPR spectrum of single molecule is detectable. Only a few data have been reported in modern surface science literature. To date it has yet to be proven to which extent ESR can be reliably and reproducibly performed on single molecules. We are setting up a new ESR-STM "spectrometer" to verify and study the effect of spin-spin correlations in the frequency spectrum of the tunneling current which flows through a single magnetic adsorbate and a metal surface. Here. we discuss, the major experimental challenges that we are attempting to overcome.**

## I. Introduction

The first experiment attempting to detect the dynamics of single spins located on a surfaces was published more than one decade ago [1]. In this experiment the authors make use of STM to study Oxidized Silicon surfaces. A magnetic field of a few hundreds Gauss was applied perpendicularly to the surface plane. By looking at the Radio Frequency (RF) tunneling current spectrum, the authors were able to locate resonances at the Larmor frequency. These resonance peaks were proved to be dependent on the magnitude of the applied magnetic field [2]. The experiment was named ESR_STM. The time dependence of this effect was also proven [3] as well as the possibility to detect the peak from spin S=1 centre [4]. Nevertheless these results were not reproduced till recently [5,6]. In these recent works the authors use drop cast radicals on HOPG as spin centre. They demonstrate that a RF resonance appears in the tunneling current spectrum at the Larmor frequency. Such a resonance is once again magnetic field dependent. These new results are particularly relevant as they point out the generality of previous measurements. Moreover they show the applicability of ESR-STM to small organic magnetic molecules. However, a number of important experimental points require a further clarification. First of all the overall amount of experimental evidences must be drastically enlarged. No previous works on this subject illustrates the RF recovery electronics in details. Particularly it is not understood if a poor capacitive matching reduces the amount of the signal delivered to the recovery electronics. Also the dependence of the RF resonance on the magnetic field requires that a much larger number of points can be acquired within a single experimental session. As an explanation of the observed ESR spectrum, it has been proposed that spin-spin correlations between tunneling electrons and the spin centre anchored at the surface can be used for extracting information on the different terms of the *single molecular spin Hamiltonian.* As spin-spin correlations act on a time scale smaller than phonon-spin interaction, the ESR spectrum of single molecules could be observed even at room temperature[7]. In order to understand the mechanism underling the observed peak a number of model parameters should be carefully tested. This requires both the ability to look at the dependence of the RF resonance on the STM tip height, bias voltage and the ability to tune the nature of the spin center as well as the easiness to locate them. We want to tune the former two molecular properties by using an approach which combines molecular electronics and molecular magnetism concepts. In particular a part of our group has been developing Self Assembled Monolayers (SAM) of organic magnetic radicals. In this paper we describe the design of an instrument that attempts to match the previous requirements.

## II. Instrument Design Considerations

The major experimental challenge to be addressed while designing an ESR-STM instrument is to maximize the Signal to Noise ratio in the RF recovery circuit. This task in the case of the STM requires both a good RF circuitry (described below) and a design that minimize thermal drift and mechanical vibrations. In fact the former two parameters affect the time the STM TIP can reliably be positioned over a surface spot of molecular size. This time is short and can vary from few hundreds milliseconds to a few seconds at room temperature. The later sets a strong difference between conventional ESR experiments and ESR_STM experiments. In fact, in this case, S/R ratio cannot be increased by repeatedly acquiring spectra over the same molecule. Also the overall amount of time available for the acquisition of a single spectra is a quite limiting factor. For these reasons we opted to design a highly symmetric , well damped STM. The entire STM is also placed in a vacuum chamber ($10^{-6}$ Torr) . Vacuum conditions ensure an improvement of the previous requested features.

The requirement to tune the magnetic properties of the adsorbates implies to use several molecules deposited in air. Samples prepared this way may result in large contaminated areas. For this reason, once the system is in vacuum conditions it is advisable to have the possibility to microscopically change the STM TIP position. We have introduced an X-Y coarse positioning apparatus to meet this requirement.

The requirement of varying the magnetic field within the course of a single experiment poses two serious problems. First of all the actual measurement of the magnetic field and second the allocation of a variable room temperature magnet in the chamber. We have solved the first problem by inserting a Hall probe close to the sample position. An electromagnet would be too cumbersome to be allocated within a small and compact STM design. It may also introduce mechanical vibrations . The resulting heating effect in the use of large currents within the magnet would generate drift in the microscope . We have inherently opted to mount a permanent magnet on to an electrically activated slide. In this way we can sweep the magnetic field applied on the sample from 100 to 4000 Gauss, by moving the magnet position.

The RF and DC recovery is the most challenging part of the entire design. From an electronic point of view the RF resonance in the tunneling current is a very low level signal ( estimated around -120 dBm) to be detected in a very small acquisition time. Also the requirement to work under different magnetic field magnitudes sets the challenge to have a broad band amplification ( from 100 MHz to 1,5 GHz). Furthermore the junction impedance is unknown. There will certainly be a number of parasitic capacitances between the STM Tip and the ground, between the connecting cable and the ground ecc.. These capacitances may drive the RF signal out of the RF recovery circuitry. Therefore parasitic capacitances should be minimized as much as possible. Fig.1 Shows the overall circuitry of the RF/DC recovery. In order to minimize the effect of parasitic capacitances we have placed an RF preamplifier close to the tunneling junction. The amplifier features 10 dB amplification and a band width of 1.2 GHz. In this way it is possible to amplifier the signal quite close to the tunnel junction. Also, once an appropriate magnetic field sub range has been chosen , it is possible to filter out the part of the RF spectrum that is not of immediate interest. to the signal out coming from the preamplifier is feed a second and a third RF commercial amplifier. The preamplifier is homemade. It is placed in vacuum, though no particular care has been taken to use vacuum compatible elements.

Final signal detection is provided by a Spectrum analyzer, as in the papers reported earlier.

RF insulation is provided by the use of highly insulated coaxial assemblies.

The requirement to work with SAM of Nytroxyl aromatic derivates sets the demand for a very high DC amplification. We have designed and constructed an high gain I/v converter according to a design published in [8 ]. In our case, the two amplification stages are placed one in vacuum and the other one outside the vacuum chamber. Further details will be published elsewhere.

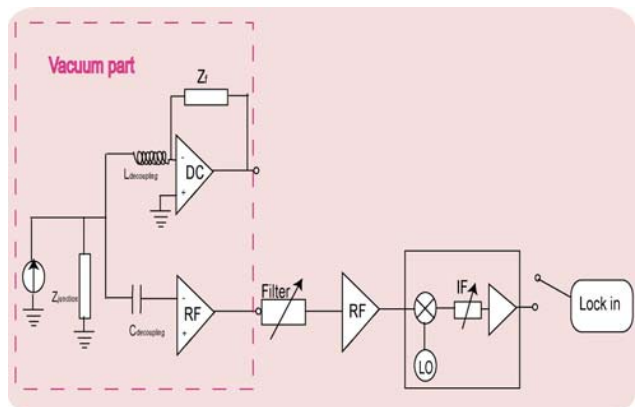

Fig. 1. Overview of the DC/RF recovery circuitry. The STM tunneling junction is considered as a current source with a parallel impedance. The current is split into the RF and DC part. The RF component is amplified close to the junction. Afterwards the signal is feed into a second amplification stage and then to a spectrum analyzer.

III. DESCRIPTION OF THE INSTRUMENT

Fig.2a,b,c, show the overall system. The vacuum chamber consists of a bottom circular pies, where the various flanges have been allocated. The top part is flanged through a N type connector. The STM is suspended on three springs. The body consists of a massive hallow cylinder and a second cylinder. A circular platform is mounted on the bottom of the hallow cylinder. This platform allocate the X-Y coarse positioning.

The movement is provided by two electrical motors. A piezo tube is mounted on top of this positioning stage. The sample is mounted on the second cylinder. This second and massive cylinder is positioned on top of the hallow cylinder. Coarse approach is provided by first manually tuning two screws and then by an electrical motor which further tunes the distance between the top cylinder and the hallow one (Tip-surface).

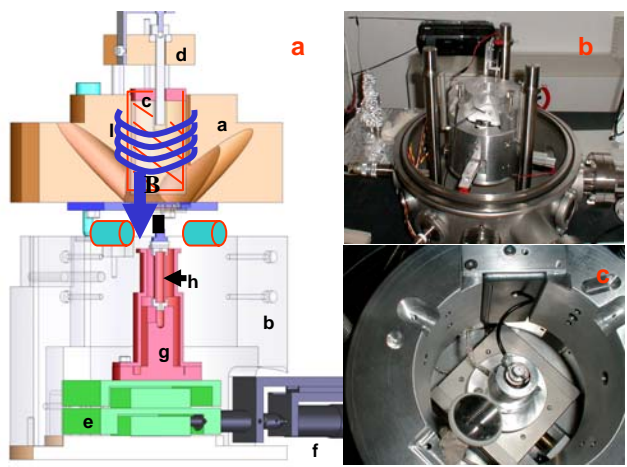

Fig. 2**a**. Drawing of the STM head: a) top cylinder, b) hallow cylinder, c) magnet, d) magnet slide, e) X-Y coarse positioning, g) piezo tube support, h) piezo tube, f) X-Y motor, l) magnet additional coil. **b** and **c** overview of the entire ESR-STM and detail of the STM tip mounted on the X-Y translation stage

The hall probe is positioned into the sample holder underneath the sample (3 mm away from the sample surface).

The piezo tube is surrounded by a special bi-cylinder piece, designed to allocate a coil. Such a coil is connected to a current amplifier and will be used for real time experiments.

The magnet is allocated in a hole provided inside the top cylinder. The hole is concentric to the piezo tube. The magnet is mounted on a slide driven by an electrical motor. The motor allows the magnet-sample distance to be varied continuously. In this way the magnitude of the magnetic field can be tuned at will, as the magnet slides up and down the hole. A coil around the magnet can provide fast variation of the magnetic field in a range of 100 gauss from the set point. The Coil is connected to a second current amplifier which receives the signal from a DAC.

The software handle both the STM standard operation and the ESR-STM Operations.

STM control electronics is standard from APEResearch.

## IV. MAGNETIC MOLECULAR SYSTEMS

The system is particularly suitable to work with SAM. In fact the high gain DC amplifier enable the scanning at low current set points. Magnetic radicals with one, two or three unpaired electrons (ferromagnetically coupled at room temperature) have been being synthesized in one of our groups. These molecules features a phenyl or bi-phenyl linker ending with an S-H or S-R tail. The tail is used for chemisorption on to Au surfaces. The second end of the linker is bound to the organic spin carrier. These are the parts of the molecule that allocate the magnetic functionality. Further information will be provided in a separate abstract.

## V. CONCLUSIONS

We have been developing an ESR-STM spectrometer. The instrument features vacuum operations, in-situ variable magnetic field, special DC and RF recovery circuitry. It is designed to specifically work with SAM and air deposited organic (inorganic) magnetic molecule. The system is currently under test. Mechanical stability and thermal drift in air have proven to be satisfactory. Further insights will be reported elsewhere.

## ACKNOWLEDGEMENT

The Authors want to thank Prof.Yshay Manassen for fruitful discussions and for the scientific support received. Financial support from the Italian MIUR, FIRB and FISR projects, and EC HPRI-CT-2000-40022 SENTINEL is also acknowledged .
* paolo.messina@unfi.it